\documentclass[aps,prl,nofootinbib,twocolumn,superscriptaddress,preprintnumbers,floatfix]{revtex4-2} 

\usepackage[nodisplayskipstretch]{setspace}
\usepackage{cancel}
\usepackage{mathtools}

\usepackage{amssymb}
\usepackage{amsmath}
\usepackage{epsfig}
\usepackage{hyperref}
\usepackage{breakurl}
\usepackage{bm}
\usepackage{color}
\usepackage{slashed}
\usepackage{tikz}
\usetikzlibrary{decorations.markings}
\usepackage[english]{babel}
\newcommand\beq{\begin{equation}}
\newcommand\eeq{\end{equation}}
\def\bea{\begin{eqnarray}}
\def\eea{\end{eqnarray}}

\DeclareRobustCommand{\SkipTocEntry}[4]{}

\newcommand{\nn}{\nonumber}

\newcommand\beal{\begin{aligned}}
\newcommand\eeal{\end{aligned}}

\usepackage{xcolor}

\newcommand{\bp}{{\boldsymbol p}}

\newcommand{\cG}{\mathcal {G}}


\begin{document}

\preprint{DESY\, 25-089\\\phantom{~}}
\title{Local-in-Time Conservative Binary Dynamics \\ [0.2cm]  at Fifth Post-Minkowskian and First Self-Force Orders} 
\author{Christoph Dlapa}
\affiliation{ Deutsches Elektronen-Synchrotron DESY, Notkestr. 85, 22607 Hamburg, Germany}

\author{Gregor K\"alin}
\affiliation{ Deutsches Elektronen-Synchrotron DESY, Notkestr. 85, 22607 Hamburg, Germany}

\author{Zhengwen Liu}
\affiliation{School of Physics and Shing-Tung Yau Center, Southeast University, Nanjing 210096, China}
\affiliation{Niels Bohr International Academy, Niels Bohr Institute, Blegdamsvej 17, 2100 Copenhagen, Denmark}

\author{Rafael A. Porto}
\affiliation{ Deutsches Elektronen-Synchrotron DESY, Notkestr. 85, 22607 Hamburg, Germany}

\begin{abstract}
We report the local-in-time conservative dynamics of nonspinning binary systems at fifth Post-Minkowskian (5PM) and first self-force (1SF) orders. This follows  from an explicit calculation of the 5PM/1SF nonlocal-in-time tail-type contribution to the deflection angle via worldline effective field theory techniques.  Proceeding as in Dlapa~{\it et al}.~\cite{4pmeftloc}, we subtract the nonlocal tail terms from the result in Driesse~{\it et al}.~\cite{Driesse:2024xad} and reconstruct a local-in-time Hamiltonian in isotropic gauge---valid for generic orbits. For completeness, we reinstate the nonlocal terms relevant for elliptic-like motion up to~6PN/1SF in a small-eccentricity expansion. Via the connection between the (source) energy flux in Dlapa~{\it et al}.~\cite{4pmeftot} and tail effects, we also derive the SF-exact logarithmic-dependent part of the full 5PM bound Hamiltonian.  Our results provide the most accurate description to date of the  dynamics of bound compact objects within the framework of relativistic scattering computations.  
\end{abstract}
\maketitle

{\bf Introduction.}  Driven by advancing our understanding of the dynamics of binary systems in General Relativity in the era of gravitational-wave (GW) astronomy~\cite{LISA,ET,CE}, perturbative frameworks to the two-body problem such as the Post-Newtonian (PN) expansion---organized as a simultaneous series in the relative velocity and Newton's constant $G$, linked through the virial theorem---and the Post-Minkowskian (PM) approximation---which resums all orders in the velocity at a fixed order in $G$---have attracted renewed interest, see e.g.~\cite{Damour:2014jta,tail,Marchand:2017pir,Foffa:2019rdf,nrgr4pn2,hered1,5pn1,5pn2,hered2,Cho:2022syn,Blanchet:2023sbv,Blumlein:2020pyo,Bini:2020wpo,binidam1,binidam2,Khalil:2022ylj,Blumlein:2021txj,memory,Almeida:2025nic,Cho:2021mqw,Amalberti:2024jaa,Bini:2024rsy,damour1,Damour:2017zjx,bohr,cheung, donal, zvi1,paper1,paper2,b2b3,Damour:2019lcq,parra,pmeft,3pmeft,tidaleft,janmogul,pmefts,Jakobsen:2021zvh,Mougiakakos:2021ckm,Gabriele2,eftrad,4pmeft,4pmeft2,4pmeftot,4pmzvi,4pmzvi2,Bini:2022enm,Jakobsen:2022zsx,Jakobsen:2023hig,Jakobsen:2023ndj,Damgaard:2023ttc,dklp,cy,Frellesvig:2023bbf,Bern:2024vqs,Bern:2024adl,Bini:2024rsy,Buonanno:2024vkx,4pmeftloc,Bini:2024tft,Driesse:2024xad,Driesse:2024feo,Bini:2025vuk}.~Recent advances in the field can be attributed, in part, to a combination of worldline effective field theory (EFT) methodologies \cite{nrgr,nrgrs,iragrg,review,Goldberger:2022ebt} with modern integration techniques from collider physics, see e.g. \cite{dklp,cy,Frellesvig:2023bbf}. The synergy has led to a rapidly evolving state of the art, which includes, on the one hand, complete results to fourth order in the PN expansion \cite{Damour:2014jta,tail,Marchand:2017pir,Foffa:2019rdf,nrgr4pn2,Cho:2022syn,Blanchet:2023sbv}, as well as potential and hereditary effects at fifth order \cite{5pn1,5pn2,hered2,Blumlein:2021txj,memory} and partial knowledge at higher orders \cite{Bini:2020wpo,binidam1,binidam2}; and on the other hand, in the realm of scattering calculations, 
the total spacetime impulse up to 4PM order \cite{4pmeft2,4pmeftot}, including also spin effects \cite{Jakobsen:2023hig,Jakobsen:2023ndj}. More recently, due to the increasing complexity of PM computations, the impulse at ${\cal O}(G^5)$ (akin of a four-loop calculation in collider physics) was tackled in combination with a self-force (SF) expansion in the mass ratio, reaching the 5PM/1SF level of accuracy for nonspinning bodies~\cite{Driesse:2024xad,Driesse:2024feo}.\vskip 4pt   

Despite these notable achievements, the PM derivations in \cite{4pmeft2,4pmeftot,Driesse:2024xad,Driesse:2024feo} all  face a major challenge when extrapolating observables from scattering to bound states. For example, the binding energy for quasi-circular orbits, obtained through the ``boundary-to-bound" (B2B) analytic continuation \cite{paper1,paper2}, fails  (other than logarithms) to match the known PN value \cite{b2b3,Khalil:2022ylj}. The discrepancy is due to nonlocal-in-time hereditary contributions~\cite{Damour:2014jta,tail}, such as tail effects arising from the scattering of the emitted radiation off of the binary's own gravitational field, which prevent the straightforward use of unbound results to describe arbitrary elliptic-like motion.\vskip 4pt This issue was addressed in \cite{4pmeftloc}, where a decomposition into local- and nonlocal-in-time effects of the full result in \cite{4pmeftot} was successfully achieved for the first time. Following the same strategy,  we present the explicit computation of nonlocal-in-time contributions to the scattering angle at 5PM/1SF order. The derivation of nonlocal tail terms involves an integral over the (source) energy spectrum times the logarithm of the center-of-mass GW frequency~\cite{b2b3,binidam2}. As discussed in \cite{4pmeftloc}, the integrand depends on the velocity and mass ratio. However, we restrict ourselves here to the 1SF order in the mass~expansion, thus reducing the integration back to a one-scale problem. (We will return to the computation of higher order terms elsewhere.) As in previous derivations \cite{dklp,4pmeftloc}, we employ the methodology of differential equations. The~nonlocal part of the 5PM/1SF deflection angle features multiple polylogarithms (MPLs) up to weight three. We find agreement in the overlap with the values in~\cite{binidam2}.\vskip 4pt  

In principle, conservative nonlocal-in-time tail-of-tail effects that are odd in the velocity also need to be considered \cite{Bini:2020wpo}. However, these were not computed by the authors of \cite{Driesse:2024xad}, who focused exclusively on even-in-velocity terms. Accordingly, we extract the universal local-in-time contribution by simply removing nonlocal tail effects from the even-in-velocity scattering angle. The corresponding center-of-mass momentum and Hamiltonian (in isotropic gauge) then follow from the framework developed in \cite{paper1,paper2,b2b3}. By leveraging the structure of the tail integrand \cite{b2b3,4pmeftloc} we derive the logarithmic-dependent part of the full 5PM Hamiltonian using the value of the (source) energy flux obtained in \cite{4pmeftot}. For completeness, we incorporate the remaining (non-logarithmic) nonlocal tail terms in a small-eccentricity approximation by adapting the calculations in \cite{binidam2} to the isotropic gauge. The combined bound Hamiltonian is fully consistent in the overlap with the state-of-the-art knowledge in the PN expansion~\cite{binidam1,binidam2,Khalil:2022ylj}, while simultaneously resumming an infinite number of (local-in-time) velocity corrections.\vskip 4pt
 {\bf Nonlocal-in-time tail effects.}  We follow the analysis in \cite{4pmeftloc} and identify nonlocal-in-time tail effects in ${\cal S}_r$, the (unbound) radial action, through its universal (gauge-invariant) structure \cite{Damour:2014jta,tail,b2b3} 
\beq
 {\cal S}^{(\rm nloc)}_{r\,} = -\frac{G E}{2\pi}   \int_{-\infty}^{+\infty} \frac{d\omega}{2\pi} \frac{dE_{\rm src}}{d\omega}  \log \left(2e^{\gamma_E}GM\omega\right)^2\,.\label{nloc1}
\eeq   
We abide by the conventions in \cite{4pmeftloc,Damour:2014jta}, with $E$ the total energy,  $\frac{dE_{\rm src}}{d\omega}$ the (odd-in-velocity) source GW spectrum in the center-of-mass frame, $M=m_1+m_2$ the total mass, and $\gamma_E$ is Euler's constant. The scattering angle is given by $\frac{\chi}{2\pi} = -\partial_J   {\cal S}_r$, where $J$ is the total angular momentum. We write the coefficients of the deflection angle in a PM expansion as, 
\beq \frac{\chi}{2} = \sum_{n=1} \left(\chi_b^{(n)} + \chi_b^{(n)\log} \log \frac{\hat b}{\Gamma}\right) \hat b^{-n},\label{eq:angle}\eeq 
where $\hat b \equiv b/GM$, with $b$ the impact parameter, $\gamma \equiv u_1\cdot u_2$, in the mostly negative metric convention, $u_a$ ($a=1,2$) are the particle's incoming velocities, and $\Gamma \equiv E/M = \sqrt{1+2\nu(\gamma-1)}$, with $\nu \equiv m_1m_2/M^2$ the symmetric mass ratio. We should stress that, although as is well known the total value for $\chi_b^{(n)}/\Gamma$ truncates in the $\nu$ expansion \cite{Damour:2019lcq,paper1}, its nonlocal-in-time contribution exhibits a much more intricate dependence on the mass ratio through the connection with the center-of-mass frame in \eqref{nloc1} \cite{4pmeftloc}. In~this letter we provide an explicit derivation of the leading term in the SF expansion at 5PM order. In contrast, using the connection between the radial action and radiated energy \cite{tail,b2b3,4pmeftot}, we are able to derive the SF-exact result for the logarithmic-dependent part.

\vskip 4pt {\bf Integrand construction.}  Our problem requires the evaluation of the integral in \eqref{nloc1} to ${\cal O}(G^4)$. Due to the dependence on the GW spectrum, we find a similar set of three-loop integrals as those introduced in \cite{4pmeftot,dklp} for the derivation of the radiated energy to 4PM, involving linear and square propagators \cite{pmeft}. The integrand features the $112$ combination, with delta functions in $\ell_{1(2)} \cdot u_1$ and $\ell_3
\cdot u_2$, and linear propagators for the remaining $\ell_{1(2)} \cdot u_2$ and $\ell_3\cdot u_1$, with $\ell_{1,2,3}$ the (three) independent loop momenta; whereas, for the quadratic propagators, we find it convenient, by means of a partial fraction decomposition, to reduce them to only one set (see the top row of Eq. (4.9) in~\cite{dklp}),
\bea
 &&\left\{ -\ell_1^2,-\ell_2^2,-\ell_3^2,-(\ell_1-q)^2,-(\ell_2-q)^2, -(\ell_3-q)^2 \right.\nn \\ 
 && \quad\,\,\,\left. -(\ell_1-\ell_2)^2,-(\ell_2-\ell_3)^2,-(\ell_1-\ell_3)^2 \right\}\,,
  \eea
  with $q^\alpha$ the momentum transfer, obeying $q\cdot u_a=0$. We~consider either retarded or advanced choices for the $i0^+$-prescription, for which only two of the propagators may be going on-shell, i.e. $k_1^\alpha=\ell^\alpha_2-\ell^\alpha_3$ or $k_2^\alpha=\ell^\alpha_1-\ell^\alpha_3$\,. In each region, we multiply by a factor of $(2e^{\gamma_E}k_{1(2)}\cdot u_{\rm com})^{2\tilde\epsilon}$  the integrand for $\frac{dE_{\rm src}}{d\omega}$, with  $u_{\rm com} \equiv \frac{(m_1 u_1+m_2u_2)}{E}$ the center-of-mass velocity. The logarithm in \eqref{nloc1} follows after expanding in  $\tilde \epsilon$, which we distinguish from the standard $D=4-2\epsilon$ in dimensional regularization. We perform a mass-ratio expansion, $m_1/m_2 \ll 1$, and retain the leading order in~$\nu$, accounting for the 1SF terms. At the end of the day, the relevant integrals are analogous to the ones in \cite{4pmeftot,dklp}, but with linear propagators that have $\tilde \epsilon$-dependent powers. Using integration-by-parts (IBP) reduction techniques implemented in \texttt{Kira}~\cite{Maierhofer:2017gsa, Klappert:2020nbg}, we find 343 master integrals for the two independent ``1rad" region(s) contributing to the source radiated energy, where each of the $k_{1(2)}$ momenta go on shell, thus isolating the relevant part of the computation.\footnote{In principle, tail effects in the impulse enter via the``2rad" region \cite{4pmeft2,4pmeftot}. However, their contribution to the nonlocal radial action factorizes into a source (rad1-only) part times the logarithm of the GW frequency \cite{b2b3}. Hence, to avoid double counting, the 2rad region must be discarded from the spectrum in~\eqref{nloc1}.}

\vskip 4pt {\bf Integration.} To solve for the master integrals, we derive differential equations in $x$, given by $\gamma=\frac{1}{2}\left(x+\frac{1}{x}\right)$. To deal with the additional $\tilde \epsilon$-dependence, it is useful to (temporarily) set $\tilde\epsilon=\eta\epsilon$, and derive canonical differential equations factorized in $\epsilon$. For this purpose, we employ standard algorithms in public packages \cite{Lee:2020zfb,Prausa:2017ltv,Meyer:2017joq,Dlapa:2020cwj}, as well as various (private) implementations which are able to deal with the extra $\eta$ parameter. Similarly to \cite{Driesse:2024xad}, we find that elliptic integrals do not feature in the final answer (since the latter appear only in the direction of the impact parameter at 4PM) and the solution depends on MPLs up to weight three, defined through the relation
\beq
\begin{aligned}
G(a_1,\ldots, a_n;z) &= \int_0^{z} \frac{dt}{t-a_1} G(a_2,\ldots,a_n;t)\,,\\\label{eq:G}
 G(\cdot ;z)&=1\,, \quad G(\underbrace{0,\ldots, 0}_{n};z) = \tfrac{1}{n!} \log^n z\,,
 \end{aligned}
 \eeq
 with $(a_n,z)$ a set of  complex numbers. We choose the variable $z=1-x$, which has a direct connection with the PN expansion, such that the result may be written in terms of a five letter {\it alphabet} $(0,1,2, 1\pm i)$. 

\vskip 4pt {\bf Scattering angle.} The derivation of the nonlocal part of the scattering angle is straightforward from the knowledge of the unbound radial action. On the one hand, for the logarithmic correction, we find
\begin{widetext}
  \begin{equation}
    \label{chinloclog}
    \begin{aligned}
      \frac{1}{\Gamma}\chi_{b (\rm nloc)}^{(5)\log} &=
      \nu \Bigg(
      \frac{\cG_1(x) h_1(x)}{90 (x-1)^6 x^3 (x+1)^6 \left(x^2+1\right)^7}
      +\frac{\cG_3(x) h_2(x)}{30 (x-1)^7 x^3 (x+1)^7 \left(x^2+1\right)^8}
      +\frac{\cG_4(x) h_3(x)}{4 (x-1)^5 x^2 (x+1)^5}\\
      &\quad\quad\,\,\,\,+\frac{\cG_6(x) h_4(x)}{(x-1)^8 x (x+1)^8}
      +\frac{\cG_7(x) h_5(x)}{2 (x-1)^5 x (x+1)^5}
      +\frac{\cG_8(x) h_6(x)}{4 (x-1)^5 x^2 (x+1)^5}
      \Bigg)\\
      &+\nu ^2 \Bigg(
      \frac{\cG_1(x) h_7(x)}{44100 (x-1)^6 x^3 (x+1)^6 \left(x^2+1\right)^8}
      +\frac{\cG_3(x) h_8(x)}{210 (x-1)^7 x^3 (x+1)^7 \left(x^2+1\right)^9}
      +\frac{\cG_4(x) h_9(x)}{4 (x-1)^3 x^2 (x+1)^5}\\
      &\quad\quad\,\,\,\,+\frac{\cG_6(x) h_{10}(x)}{(x-1)^8 x^2 (x+1)^8}
      +\frac{\cG_7(x) h_{11}(x)}{4 (x-1)^5 x^2 (x+1)^5}
      +\frac{\cG_8(x) h_{12}(x)}{4 (x-1)^5 x^2 (x+1)^5}
      \Bigg) \,,
    \end{aligned}
  \end{equation}
\end{widetext}
which we upgraded to its SF-exact expression, as we mentioned, thanks to the relationship between tail terms and radiated energy \cite{dklp,4pmeftot}. For the remaining non-logarithmic terms, on the other hand, we have
\begin{widetext}
  \beq\label{chinlocnlog}
  \begin{aligned}
    \frac{\chi_{(\rm nloc)}^{(5)(\rm 1SF)}}{\Gamma \nu}  &=
    \frac{\cG_1(x) h_{13}(x)}{25200 (x-1)^6 x^3 (x+1)^6 \left(x^2+1\right)^7}
    +\frac{\cG_2(x) h_{14}(x)}{90 (x-1)^6 x^3 (x+1)^6 \left(x^2+1\right)^7}
    +\frac{\cG_3(x) h_{15}(x)}{12600 (x-1)^7 x^3 (x+1)^7 \left(x^2+1\right)^8}\\
    &+\frac{\cG_4(x) h_{16}(x)}{1680 (x-1)^7 x^5 (x+1)^7}
    +\frac{\cG_5(x) h_{17}(x)}{30 (x-1)^7 x^3 (x+1)^7 \left(x^2+1\right)^8}
    +\frac{\cG_6(x) h_{18}(x)}{420 (x-1)^8 x^3 (x+1)^8}\\
    &+\frac{\cG_7(x) h_{19}(x)}{336 (x-1)^5 x^5 (x+1)^5 \left(x^2+1\right)^8}
    +\frac{\cG_8(x) h_{20}(x)}{1680 (x-1)^7 x^5 (x+1)^7}
    +\frac{\cG_{10}(x) h_{21}(x)}{4 (x-1)^5 x^2 (x+1)^5}
    +\frac{\cG_{11}(x) h_{22}(x)}{2 (x-1)^5 x (x+1)^5}\\
    &+\frac{\cG_{12}(x) h_{23}(x)}{4 (x-1)^5 x^2 (x+1)^5}
    +\frac{\cG_{13}(x) h_{24}(x)}{4 (x-1)^8 x^2 (x+1)^8}
    +\frac{\cG_{14}(x) h_{25}(x)}{(x-1)^8 x (x+1)^8}
    +\frac{\cG_{15}(x) h_{26}(x)}{(x-1)^8 x (x+1)^8}\\
    &+\frac{\cG_{16}(x) h_{27}(x)}{(x-1)^5 x (x+1)^5}
    +\frac{\cG_{17}(x) h_{28}(x)}{4 (x-1)^8 x^2(x+1)^8}
    +\frac{\cG_{19}(x) h_{29}(x)}{2 (x-1)^5 x (x+1)^5}
    +\frac{\cG_{20}(x) h_5(x)}{2 (x-1)^5 x (x+1)^5}\\
    &+\frac{\cG_{22}(x) h_{30}(x)}{4 (x-1)^5 x^2 (x+1)^5}
    +\frac{\cG_{23}(x) h_{31}(x)}{4 (x-1)^5 x^2 (x+1)^5}
    +\frac{\cG_{24}(x) h_{22}(x)}{2 (x-1)^5 x (x+1)^5}\,.
  \end{aligned}
  \eeq
\end{widetext}
In both these equations the $h_i$'s are polynomials in $x$, up to degree 38, whereas the $\cG_i$'s are products of the MPLs introduced in \eqref{eq:G} up to weight three. See the supplemental material for explicit expressions.\vskip 4pt 

As emphasized in \cite{4pmeftloc}, the definition of nonlocal-in-time terms in \cite{binidam1,binidam2} includes in addition to \eqref{nloc1} ($W_1$ in \cite{binidam2}) an extra term involving an integral over the GW flux times $\log r(t)$. Since the latter is coordinate dependent, and generated by (quasi-instantaneous) potential-only interactions, we keep it instead as part of the local-in-time side of the dynamics. We~have checked that the results in \eqref{chinloclog} and \eqref{chinlocnlog} nicely agree with the $W_1$-only PN values of the scattering angle at ${\cal O}(G^5)$ computed in \cite{binidam2}.\vskip 4pt  Upon subtracting nonlocal terms from the even-in-velocity scattering angle in \cite{Driesse:2024xad}, $\chi_{b \rm (even)}^{(5)\rm 1SF}$, we arrive~at
\beq
\chi_{b (\rm loc)}^{(5) (1\rm SF)} = \chi_{b \rm (even)}^{(5)  (1\rm SF)} - \chi_{b (\rm nloc)}^{(5)}\,,\quad
\chi_{b (\rm loc)}^{(5)\log} = - \chi_{b (\rm nloc)}^{(5) \log}\,,\label{chiloc}
\eeq 
for the local-in-time contribution at 5PM/1SF order. The result in \eqref{chiloc} can now be used to describe generic bound orbits as we discuss momentarily. \vskip 4pt 
{\bf Local-in-time dynamics.} The~(gauge-invariant) local-in-time bound radial action follows directly from the scattering angle \cite{paper1,paper2}. However, the B2B map implies that  odd PM orders of the angle cancel out upon analytic continuation. Nevertheless, 
 from the knowledge of the local-in-time angle we can readily obtain the center-of-mass momentum  ($\hat \bp \equiv \bp/(M\nu)$) in isotropic gauge,
\beq
{\hat \bp}^2 = \frac{v_{\infty}^2}{\Gamma^2} \left( 1 + \sum_{n=1} \frac{1}{\hat r^n} \left( f_{n}  + f_{n}^{\log} \log \hat r \right)\right) \,,
\eeq
where $v_\infty^2 \equiv \gamma^2-1$, and a Hamiltonian ($\hat H \equiv H/(M\nu)$),
\beq
\label{H}
\hat H =  \hat E_0+ \sum_{n=1} \frac{1}{\hat r^n} \left( \hat c_{n}+ \hat c_{n}^{\log}  \log \hat r\right)\,,
\eeq
with $\hat r \equiv  \frac{r}{GM}$,  $\hat E_0 = \frac{1}{M\nu} \sum_a E_a$, and $E_a \equiv \sqrt{\bp^2+m_a^2}$.\vskip 4pt

{\bf Universal logarithms.} 
 The coefficient of the $\log r$ term in the total bound Hamiltonian at ${\cal O}(G^n)$ is related to the (source) energy flux at ${\cal O}(G^{(n-1)})$. This follows from the optical theorem in conjunction with the cancellation of ultraviolet and infrared divergences (alongside factors of $\log b$) from local- and nonlocal-in-time effects within dimensional regularization \cite{tail,b2b3,4pmeft,4pmeftloc}.  This implies, for the logarithmic coefficient of the 5PM Hamiltonian for elliptic-like motion, the relationship
\bea
\hat c_{5(\rm tot)}^{\rm ell, \log} \label{Hlog} 
&=&  -G\left. \left( \hat H \frac{dE_{\rm src}}{dt}\right)\right|_{ G^4}\\\,
&=& -G \left(\left. \frac{\hat c_1}{\hat r}\frac{dE_{\rm src}}{dt}\right|_{ G^3}+ \left.\hat E_0 \frac{dE_{\rm src}}{dt}\right|_{ G^4}\right) \,. \nn
\eea
The source energy flux was given in \cite{4pmeft,4pmeftloc} at ${\cal O}(G^3)$, whereas the 4PM component (including only the ``rad1" piece of the radiated energy) reads  \cite{4pmeftot,b2b3}
\begin{widetext}
\bea
  \begin{aligned}
  G \left.\frac{dE_{\rm src}}{dt}\right|_{ G^4} (\hat r,\bp^2) &= \frac{\nu ^3}{\Gamma ^3 x^4 \left(x^2-1\right)^2 \xi \hat{r}^5}
  \left[
    \frac{\log (2) d_1(x)}{256 \Gamma ^7 x^4 \xi ^2}
    +\frac{d_2(x) \log ^2(x)}{\left(x^2-1\right)^4}
    +\frac{d_3(x) \text{Li}_2(x)}{256 \left(x^2-1\right)}
    +\frac{d_4(x) \text{Li}_2(-x)}{256 \left(x^2-1\right)}\right.\\
    &\quad+\frac{d_6(x) \log (x+1)}{256 \Gamma ^7 x^4 \xi ^2}
    +\frac{d_7(x) \log ^2\left(x^2+1\right)}{1024 \left(x^2-1\right)}
    +\frac{d_7(x) \text{Li}_2\left(\frac{1}{x^2+1}\right)}{512 \left(x^2-1\right)}
    +\frac{\pi ^2 d_8(x)}{2048 \left(x^2-1\right)}\\
    &\quad+\frac{d_9(x)}{15052800 \Gamma ^9 x^5 \left(x^2-1\right)^2 \left(x^2+1\right)^8 \xi ^2}
    +\log (x) \left(
      \frac{d_3(x) \log (1-x)}{256 \left(x^2-1\right)}
      +\frac{d_4(x) \log (x+1)}{256 \left(x^2-1\right)}\right.\\
      &\quad\left.\left.+\frac{d_5(x)}{71680 \Gamma ^9 x^5 \left(x^2-1\right)^3 \left(x^2+1\right)^9 \xi ^2}\right)
  \right]\,,\label{dE4}
  \end{aligned}
\eea
\end{widetext}
where $\xi \equiv \frac{E_1E_2}{(E_1+E_2)^2}$, $x= \gamma - \sqrt{\gamma^2-1}$, and $\gamma = (E_1E_2+\bp^2)/m_1m_2$. The $d_i$'s are polynomials in $x$ up to degree 38, that also depend on $\Gamma$ and $\nu$. See supplemental material.\vskip 4pt 

{\bf Assembling the pieces.}  Collecting all the ingredients, the bound Hamiltonian to 5PM order takes the form
\bea
\label{Htot5pm} \hat H^{\rm ell}_{\rm 5PM}  &=&  \hat E_0+\sum_{i=1}^{i=5} \frac{\hat c_{i\rm (loc)}}{\hat r^i}  + \sum_{i=1}^{i=5}  \frac{\hat c_{i\rm (nloc)}}{\hat r^i}  \\ &-& \sum_{i=4}^{i=5}   \left. G \left(\hat H\frac{dE_{\rm src}}{dt}\right)\right|_{ (i-1)\rm PM} \log \left(\frac{\hat r}{e^{2\gamma_E}}\right) \,,  \nn 
 \eea
after absorbing the $e^{2\gamma_E}$ from \eqref{nloc1} into the logarithm. The $\hat c_{1|2|3|(\rm loc)}$ are the PM coefficients up to ${\cal O}(G^3)$  computed in \cite{cheung,pmeft,zvi1,3pmeft}, the $\hat c_{4(\rm loc)}$ was first obtained in \cite{4pmeftloc}, while the (flux-dependent) SF-exact coefficient of the logarithm follows from the results in \cite{4pmeftot}. The value for $\hat c_{5(\rm loc)}$ is reported here at 1SF order for the first time.\vskip 4pt The remaining \big[non-$\log \left(\frac{\hat r}{ e^{2\gamma_E}}\right)$\big] $\hat c_{i\rm (nloc)}$ terms for elliptic-like motion, however, are not known in a PM scheme. Nevertheless, similar coefficients (in another gauge) have been computed within the framework of the PN approximation in a small-eccentricity expansion. Adapting the  results in~\cite{binidam2} to our isotropic gauge, we include their values at ${\cal O}(G^5)$ up to 6PN order (see the supplemental material).~After incorporating the results in \cite{4pmeftloc}, the  expression in \eqref{Htot5pm} is perfectly consistent in the overlap with the state-of-the-art Hamiltonian inferred from the PN results in \cite{binidam1,binidam2}, truncated to 1SF order.\vskip 4pt

\vskip 4pt {\bf Conclusions.} Recent developments in integration techniques, combined with the worldline EFT framework, have substantially advanced our understanding of scattering processes in General Relativity---capturing both conservative and dissipative effects \cite{4pmeft2,4pmeftot,Driesse:2024xad,Driesse:2024feo}. However, as shown in \cite{b2b3,Khalil:2022ylj}, although local-in-time components and logarithmic terms are universal, the full hyperbolic results fail to accurately describe binaries in low-eccentricity orbits due to (orbit-dependent) nonlocal-in-time effects. This limitation was successfully addressed in \cite{4pmeftloc}, where a separation between local and nonlocal tail effects in the confines of the PM expansion was first established  at ${\cal O}(G^4)$.\vskip 4pt Following the strategy in~\cite{4pmeftloc}, in this letter we computed the tail-type nonlocal-in-time contribution to the conservative scattering angle at 5PM/1SF order, and subtracted it from the result in \cite{Driesse:2024xad} thus isolating its local-in-time component. Using the framework developed in \cite{paper1,paper2}, we derived expressions for the local-in-time center-of-mass momentum and Hamiltonian in isotropic gauge, including the SF-exact logarithmic contribution, all valid for arbitrary bound orbits. By incorporating nonlocal-in-time (non-logarithmic) tail effects for elliptic-like trajectories in the PN expansion \cite{binidam2}, we constructed a hybrid Hamiltonian that presently provides the most accurate description of binary compact objects derived from PM/PN data. Ready-to-use expressions for all relevant quantities, including also PN-expanded values to 30th order, are collected in the ancillary files.\vskip 4pt Beyond its natural relevance for GW physics, the results presented here
illustrate the remarkable power of the worldline EFT approach \cite{nrgr,nrgrs,iragrg,review,Goldberger:2022ebt} combined with modern tools, such as differential equations and IBP techniques, to tackle the integration problem \cite{dklp,cy,Frellesvig:2023bbf}. These methods offer a robust framework for addressing the full two-body dynamics in General Relativity within a perturbative scheme. We are currently exploring the extension of these results to higher PM/SF orders and the direct implementation of nonlocal-in-time effects for bound systems in the PM expansion.\vskip 4pt 

{\bf Acknowledgements}. We thank Johann Usovitsch for his help with using \texttt{Kira}. We thank Donato Bini and Mohammed Khalil for their help in adapting the results in \cite{binidam1,binidam2} to the isotropic gauge. The work of CD, GK, and RAP was supported in part by the ERC-CoG Precision Gravity: From the LHC to LISA, provided by the European Research Council (ERC) under the European Union's H2020 research and innovation programme (grant No. 817791). ZL was supported partially by the European Union's Horizon Europe research and innovation program under the Marie Sk\l{}odowska-Curie grant agreement No 101146918 `{GWtheory}' and by the Start-up Research Fund of Southeast University No.\,RF1028624160.

\bibliographystyle{apsrev4-1}
 \bibliography{ref4PM}
 
 \newpage
 \begin{widetext}

 \section{Supplemental Material}
 As we did in \cite{4pmeftloc}, a hybrid formulation for elliptic-like orbits can be developed by incorporating into the Hamiltonian in~\eqref{Htot5pm} the nonlocal-in-time contributions at ${\cal O}(G^5)$---specifically the ``$W_1$-only" terms excluding the $\log\left(\frac{\hat r}{e^{2\gamma_E}}\right)$ dependence---computed up to 6PN and ${\cal O}(e^8)$ in the small-eccentricity expansion in~\cite{binidam2}, translated to our isotropic gauge and truncated to 1SF order.\footnote{We are grateful to Donato Bini and Mohammed Khalil for their help computing these (unpublished) values.} The total bound Hamiltonian then takes the  form,
 \bea
\label{Htot}\hat H^{\rm ell}_{{\rm hyb}} &=& \hat E_0 + \sum_{i=1}^{i=5} \frac{\hat c_{i\rm (loc)}}{\hat r^i}  - \sum_{i=4}^{i=5}   \left. G \left(\hat H \frac{dE_{\rm src}}{dt}\right)\right|_{G^{(i-1)}} \log \left(\frac{\hat r}{e^{2\gamma_E}}\right)+ \sum_{i=1}^{i=5}  \frac{1}{\hat r^i}  \left\{\hat c^{{\rm 6PN}(e^8)}_{i\rm (nloc)}+ {\cal O}\big(\hat\bp^{2(8-i)}\big)\right\}\,,
\eea
with the $c^{{\rm 6PN}(e^8)}_{i(\rm nloc)}$ ($i \leq 4$) given in \cite{4pmeftloc}, and 
\bea
\begin{aligned}
  \frac{\hat c^{{\rm 6PN}(e^8, {\rm 1SF})}_{5(\rm nloc)}}{\nu}
  &= \bigg(-383 + \frac{5421492}{5}\log(2) + \frac{631071}{20}\log (3)-\frac{1953125}{4} \log(5) \bigg)\\
  &\quad+\hat\bp^2 \begin{multlined}[t]
    \bigg(-\frac{237374547}{28000}+\frac{274846146629 \log (2)}{9450}+\frac{268234748343 \log (3)}{71680}-\frac{5343159203125 \log (5)}{387072}\\
    -\frac{96889010407 \log (7)}{92160}\bigg)
  \end{multlined}\\
  &\quad+\hat\bp^4 \begin{multlined}[t]
    \bigg(-\frac{12514611952561}{197568000}+\frac{345079750666673 \log (2)}{1190700}+\frac{2379761187413283 \log (3)}{40140800}\\
    -\frac{7512099894456875 \log (5)}{55738368}-\frac{334452591435529 \log (7)}{13271040}\bigg)\,.
  \end{multlined}
\end{aligned}
\eea
The Hamiltonian in \eqref{Htot} is then in perfect agreement in the overlapping realm of validity with the results in \cite{binidam1,binidam2}, while resuming an infinite tower of (local-in-time) velocity corrections. The explicit expressions for the $\cG_i$'s entering in \eqref{chinloclog} and \eqref{chinlocnlog}, as well as the $h_i$ polynomials, are given below, together with the $d_i$ coefficients entering in \eqref{dE4}. Ready-to-use expressions for all the relevant quantities reported in this work are collected in the ancillary files.

\begingroup
\allowdisplaybreaks
\begin{align}
  \cG_1(x) &= 1\,,\nn\\
  \cG_2(x) &= G(0;1-x)+G(2;1-x)\,,\nn\\
  \cG_3(x) &= G(1;1-x)\,,\nn\\
  \cG_4(x) &= G(0,1;1-x)\,,\nn\\
  \cG_5(x) &= G(0;1-x) G(1;1-x)-G(0,1;1-x)+G(1,2;1-x)\,,\nn\\
  \cG_6(x) &= \frac{1}{2} G(1;1-x)^2\,,\nn\\
  \cG_7(x) &= G(1-i,1;1-x)+G(1+i,1;1-x)\,,\nn\\
  \cG_8(x) &= G(1;1-x) G(2;1-x)-G(1,2;1-x)\,,\nn\\
  \cG_9(x) &= G(0,0,1;1-x)+\frac{1}{2} G(0,1,1;1-x)\,,\nn\\
  \cG_{10}(x) &= -\frac{1}{2} G(2;1-x) G(1;1-x)^2+G(1,2;1-x) G(1;1-x)+G(0;1-x) G(0,1;1-x)-2 G(0,0,1;1-x)\nn\\
  &\quad-2 G(0,1,1;1-x)+G(0,1,2;1-x)-G(1,1,2;1-x)\,,\nn\\
  \cG_{11}(x) &= -G(0,1,1;1-x)+G(0,1-i,1;1-x)+G(0,1+i,1;1-x)\,,\nn\\
  \cG_{12}(x) &= \frac{1}{2} G(2;1-x) G(1;1-x)^2-G(1,2;1-x) G(1;1-x)-\frac{1}{2} G(0,1,1;1-x)+G(0,2,1;1-x)\nn\\
  &\quad+G(1,1,2;1-x)\,,\nn\\
  \cG_{13}(x) &= G(1;1-x) G(0,1;1-x)\,,\nn\\
  \cG_{14}(x) &= \frac{1}{6} G(1;1-x)^3\,,\nn\\
  \cG_{15}(x) &= \frac{1}{2} G(0;1-x) G(1;1-x)^2-G(2;1-x) G(1;1-x)^2-(G(0,1;1-x)-2 G(1,2;1-x)) G(1;1-x)\nn\\
  &\quad-G(0,1,1;1-x)-G(1,1,2;1-x)\,,\nn\\
  \cG_{16}(x) &= G(1,1-i,1;1-x)+G(1,1+i,1;1-x)\,,\\
  \cG_{17}(x) &= G(1;1-x) (G(1;1-x) G(2;1-x)-G(1,2;1-x))\,,\nn\\
  \cG_{18}(x) &= \frac{1}{2} \left[G(2;1-x) G(1;1-x)^2+(-2 G(1,2;1-x)+G(1-i,1;1-x)+G(1+i,1;1-x)) G(1;1-x)\right.\nn\\
    &\quad\quad\,\,\left.+2 G(0,1,1;1-x)+2 G(1,1,2;1-x)-G(1,1-i,1;1-x)-G(1,1+i,1;1-x)\right]\,,\nn\\
  \cG_{19}(x) &= -\frac{1}{2} G(2;1-x) G(1;1-x)^2+G(1,2;1-x) G(1;1-x)+G(0;1-x)  (G(1-i,1;1-x)+G(1+i,1;1-x))\nn\\
  &\quad-G(0,1,1;1-x)-G(0,1-i,1;1-x)-G(0,1+i,1;1-x)-G(1,1,2;1-x)-G(1-i,0,1;1-x)\nn\\
  &\quad+G(1-i,1,2;1-x)-G(1+i,0,1;1-x)+G(1+i,1,2;1-x)\,,\nn\\
 \cG_{20}(x) &= \frac{1}{2} G(2;1-x) G(1;1-x)^2-G(1,2;1-x) G(1;1-x)+G(0,1,1;1-x)+G(1,1,2;1-x)\nn\\
  &\quad+G(1-i,1-i,1;1-x)+G(1-i,1+i,1;1-x)+G(1+i,1-i,1;1-x)+G(1+i,1+i,1;1-x)\,,\nn\\
  \cG_{21}(x) &= \frac{3}{2} G(2;1-x) G(1;1-x)^2-3 G(1,2;1-x) G(1;1-x)+3 G(0,1,1;1-x)+3 G(1,1,2;1-x)\nn\\
  &\quad+G(1-i,0,1;1-x)+G(1-i,2,1;1-x)+G(1+i,0,1;1-x)+G(1+i,2,1;1-x)\,,\nn\\
  \cG_{22}(x) &= G(2;1-x) G(0,1;1-x)+G(0,1,1;1-x)-G(0,1,2;1-x)-G(0,2,1;1-x)\nn\\
  &\quad+\frac{1}{4} \left[-G(2;1-x) G(1;1-x)^2+2 G(1,2;1-x) G(1;1-x)-2 G(1,1,2;1-x)\right]\,,\nn\\
  \cG_{23}(x) &= -G(2;1-x) G(1;1-x)^2+2 G(1,2;1-x) G(1;1-x)-G(2;1-x) G(0,1;1-x)\nn\\
  &\quad+G(0;1-x) (G(1;1-x) G(2;1-x)-G(1,2;1-x))+G(2;1-x) G(1,2;1-x)-G(0,1,1;1-x)\nn\\
  &\quad+G(0,1,2;1-x)-2 G(1,1,2;1-x)-2 G(1,2,2;1-x)\,,\nn\\
  \cG_{24}(x) &= -\frac{1}{2} G(2;1-x) G(1;1-x)^2+G(1,2;1-x) G(1;1-x)\nn\\
  &\quad+G(2;1-x) (G(1-i,1;1-x)+G(1+i,1;1-x))-G(1,1,2;1-x)-G(1-i,1,2;1-x)\nn\\
  &\quad-G(1-i,2,1;1-x)-G(1+i,1,2;1-x)-G(1+i,2,1;1-x)\,,\nn\\
  \cG_{25}(x) &= \frac{1}{4} \left[G(2;1-x) G(1;1-x)^2+2 \left(G(2;1-x)^2-G(1,2;1-x)\right)  G(1;1-x)\right.\nn\\
    &\quad\quad\,\,\left.-4 G(2;1-x) G(1,2;1-x)+2 G(1,1,2;1-x)+4 G(1,2,2;1-x)\right]\nn
\end{align}

\begin{align}
  h_1(x) &= -192 x^{32}+70345 x^{30}-1012248 x^{28}-16796521 x^{26}-53924064 x^{24}+26867973 x^{22}+515430360 x^{20}\nn\\
  &\quad+1341361563 x^{18}+1778793408 x^{16}+1341361563 x^{14}+515430360 x^{12}+26867973 x^{10}-53924064 x^8\nn\\
  &\quad-16796521 x^6-1012248 x^4+70345 x^2-192\,,\nn\\
  h_2(x) &= -768 x^{36}-40509 x^{34}-29793 x^{32}+7333086 x^{30}+47004293 x^{28}+67036080 x^{26}-297881161 x^{24}\nn\\
  &\quad-1547420350 x^{22}-3342011611 x^{20}-4243955494 x^{18}-3342011611 x^{16}-1547420350 x^{14}-297881161 x^{12}\nn\\
  &\quad+67036080 x^{10}+47004293 x^8+7333086 x^6-29793 x^4-40509 x^2-768\,,\nn\\
  h_3(x) &= 525 x^{14}-2498 x^{13}+5430 x^{12}-6324 x^{11}-5888 x^{10}+30882 x^9-60483 x^8+70568 x^7-60483 x^6\nn\\
  &\quad+30882 x^5-5888 x^4-6324 x^3+5430 x^2-2498 x+525\,,\nn\\
  h_4(x) &= 512 \left(7 x^{18}-675 x^{14}+189 x^{12}+4833 x^{10}+3159 x^8-5 x^6-213 x^4+1\right)\,,\nn\\
  h_5(x) &= -1823 x^{12}-6054 x^{10}+202671 x^8+544300 x^6+202671 x^4-6054 x^2-1823\,,\nn\\
  h_6(x) &= -525 x^{14}-2498 x^{13}-5430 x^{12}-6324 x^{11}+5888 x^{10}+30882 x^9+60483 x^8+70568 x^7+60483 x^6\nn\\
  &\quad+30882 x^5+5888 x^4-6324 x^3-5430 x^2-2498 x-525\,,\nn\\
  h_7(x) &= 188160 x^{34}-39829184 x^{33}-68749940 x^{32}-358154923 x^{31}+923064940 x^{30}+6460709986 x^{29}\nn\\
  &\quad+17452593620 x^{28}+58553910415 x^{27}+69306173300 x^{26}+163660060004 x^{25}+26514969180 x^{24}\nn\\
  &\quad+64744381653 x^{23}-531452366340 x^{22}-695910106626 x^{21}-1819656084540 x^{20}-1960277378745 x^{19}\nn\\
  &\quad-3057751871580 x^{18}-2626177239560 x^{17}-3057751871580 x^{16}-1960277378745 x^{15}\nn\\
  &\quad-1819656084540 x^{14}-695910106626 x^{13}-531452366340 x^{12}+64744381653 x^{11}+26514969180 x^{10}\nn\\
  &\quad+163660060004 x^9+69306173300 x^8+58553910415 x^7+17452593620 x^6+6460709986 x^5\nn\\
  &\quad+923064940 x^4-358154923 x^3-68749940 x^2-39829184 x+188160\,,\nn\\
  h_8(x) &= 10752 x^{38}+170752 x^{37}+577878 x^{36}+3053105 x^{35}+984228 x^{34}-23188558 x^{33}-102246102 x^{32}\nn\\
  &\quad-395889516 x^{31}-760723306 x^{30}-1705980266 x^{29}-1596565222 x^{28}-2276395052 x^{27}+3231831134 x^{26}\nn\\
  &\quad+5854097682 x^{25}+25834221154 x^{24}+31361563244 x^{23}+68452047454 x^{22}+66134059430 x^{21}\nn\\
  &\quad+106203539470 x^{20}+83073443318 x^{19}+106203539470 x^{18}+66134059430 x^{17}+68452047454 x^{16}\nn\\
  &\quad+31361563244 x^{15}+25834221154 x^{14}+5854097682 x^{13}+3231831134 x^{12}-2276395052 x^{11}\nn\\
  &\quad-1596565222 x^{10}-1705980266 x^9-760723306 x^8-395889516 x^7-102246102 x^6-23188558 x^5\nn\\
  &\quad+984228 x^4+3053105 x^3+577878 x^2+170752 x+10752\,,\nn\\
  h_9(x) &= -\left(\left(x^2+1\right) \left(x^4-4 x^2+1\right) \left(525 x^6-3046 x^5+6403 x^4-8724 x^3+6403 x^2-3046 x+525\right)\right)\,,\nn\\
  h_{10}(x) &= -512 \left(x^{20}+14 x^{19}+10 x^{18}-165 x^{16}-1350 x^{15}-772 x^{14}+378 x^{13}+2744 x^{12}+9666 x^{11}+6572 x^{10}\right.\nn\\
  &\quad\left.+6318 x^9+2756 x^8-10 x^7-700 x^6-426 x^5-217 x^4+10 x^2+2 x+1\right)\,,\nn\\
  h_{11}(x) &= -525 x^{14}+7292 x^{13}+585 x^{12}+24216 x^{11}+54643 x^{10}-810684 x^9+90705 x^8-2177200 x^7+90705 x^6\nn\\
  &\quad-810684 x^5+54643 x^4+24216 x^3+585 x^2+7292 x-525\,,\nn\\
  h_{12}(x) &= 1575 x^{14}+5896 x^{13}+10275 x^{12}+13008 x^{11}-13171 x^{10}-62088 x^9-125655 x^8-143008 x^7-125655 x^6\nn\\
  &\quad-62088 x^5-13171 x^4+13008 x^3+10275 x^2+5896 x+1575\,,\nn\\
  h_{13}(x) &= 951168 x^{32}+45900859 x^{30}+51945775 x^{28}-3546200611 x^{26}-24244274722 x^{24}-78200280441 x^{22}\nn\\
  &\quad-156383651615 x^{20}-223556226207 x^{18}-247699144412 x^{16}-223556226207 x^{14}-156383651615 x^{12}\nn\\
  &\quad-78200280441 x^{10}-24244274722 x^8-3546200611 x^6+51945775 x^4+45900859 x^2+951168\nn\\
  &\quad+560 \left(192 x^{32}-70345 x^{30}+1012248 x^{28}+16796521 x^{26}+53924064 x^{24}-26867973 x^{22}\right.\nn\\
  &\quad-515430360 x^{20}-1341361563 x^{18}-1778793408 x^{16}-1341361563 x^{14}-515430360 x^{12}-26867973 x^{10}\nn\\
  &\quad\left.+53924064 x^8+16796521 x^6+1012248 x^4-70345 x^2+192\right) \log (2)\,,\nn\\
  h_{14}(x) &= 192 x^{32}-70345 x^{30}+1012248 x^{28}+16796521 x^{26}+53924064 x^{24}-26867973 x^{22}-515430360 x^{20}\nn\\
  &\quad-1341361563 x^{18}-1778793408 x^{16}-1341361563 x^{14}-515430360 x^{12}-26867973 x^{10}+53924064 x^8\nn\\
  &\quad+16796521 x^6+1012248 x^4-70345 x^2+192\,,\nn\\
  h_{15}(x) &= -204672 x^{36}-696516 x^{34}-351764697 x^{32}-5868328356 x^{30}-21855451003 x^{28}+28879426080 x^{26}\nn\\
  &\quad+423089614151 x^{24}+1405082399700 x^{22}+2571576723341 x^{20}+2971386471024 x^{18}+2217835069901 x^{16}\nn\\
  &\quad+1037024194500 x^{14}+263670375431 x^{12}+16653367760 x^{10}-7040142523 x^8-1145605876 x^6\\
  &\quad-68389017 x^4-20393116 x^2-150912+840 \left(768 x^{36}+40509 x^{34}+29793 x^{32}-7333086 x^{30}\right.\nn\\
  &\quad-47004293 x^{28}-67036080 x^{26}+297881161 x^{24}+1547420350 x^{22}+3342011611 x^{20}+4243955494 x^{18}\nn\\
  &\quad+3342011611 x^{16}+1547420350 x^{14}+297881161 x^{12}-67036080 x^{10}-47004293 x^8-7333086 x^6\nn\\
  &\quad\left.+29793 x^4+40509 x^2+768\right) \log (2)\,,\nn\\
  h_{16}(x) &= -15360 x^{24}+119808 x^{22}-185850 x^{21}+4669125 x^{20}-3106950 x^{19}-18036936 x^{18}+31412980 x^{17}\nn\\
  &\quad-240496596 x^{16}+168688780 x^{15}+448451848 x^{14}-196808960 x^{13}+1186429342 x^{12}-196808960 x^{11}\nn\\
  &\quad+448451848 x^{10}+168688780 x^9-240496596 x^8+31412980 x^7-18036936 x^6-3106950 x^5\nn\\
  &\quad+4669125 x^4-185850 x^3+119808 x^2-15360-840 \left(x^2-1\right)^2 \left(525 x^{14}-2498 x^{13}+5430 x^{12}-6324 x^{11}\right.\nn\\
  &\quad-5888 x^{10}+30882 x^9-60483 x^8+70568 x^7-60483 x^6+30882 x^5-5888 x^4-6324 x^3+5430 x^2\nn\\
  &\quad\left.-2498 x+525\right) x^3 \log (2)\,,\nn\\
  h_{17}(x) &= 768 x^{36}+40509 x^{34}+29793 x^{32}-7333086 x^{30}-47004293 x^{28}-67036080 x^{26}+297881161 x^{24}\nn\\
  &\quad+1547420350 x^{22}+3342011611 x^{20}+4243955494 x^{18}+3342011611 x^{16}+1547420350 x^{14}\nn\\
  &\quad+297881161 x^{12}-67036080 x^{10}-47004293 x^8-7333086 x^6+29793 x^4+40509 x^2+768\,,\nn\\
  h_{18}(x) &= 23040 x^{24}-148992 x^{22}-4639573 x^{20}+25857531 x^{18}+586492950 x^{16}-726118970 x^{14}\nn\\
  &\quad-3828293052 x^{12}-1573668348 x^{10}+318087434 x^8+19761210 x^6-8862735 x^4+698369 x^2+32256\nn\\
  &\quad-430080 \left(7 x^{20}-675 x^{16}+189 x^{14}+4833 x^{12}+3159 x^{10}-5 x^8-213 x^6+x^2\right) \log (2)\,,\nn\\
  h_{19}(x) &= -6144 x^{36}-30720 x^{34}+898179 x^{32}+4285158 x^{30}-127956687 x^{28}-1319480036 x^{26}-5889432997 x^{24}\nn\\
  &\quad-15595791494 x^{22}-27197011983 x^{20}-32592942840 x^{18}-27197011983 x^{16}-15595791494 x^{14}\nn\\
  &\quad-5889432997 x^{12}-1319480036 x^{10}-127956687 x^8+4285158 x^6+898179 x^4-30720 x^2-6144\nn\\
  &\quad+336 \left(x^2+1\right)^8 \left(1823 x^{12}+6054 x^{10}-202671 x^8-544300 x^6-202671 x^4+6054 x^2+1823\right) x^4 \log (2)\,,\nn\\
  h_{20}(x) &= -15360 x^{24}+119808 x^{22}+185850 x^{21}+4669125 x^{20}+3106950 x^{19}-18036936 x^{18}-31412980 x^{17}\nn\\
  &\quad-240496596 x^{16}-168688780 x^{15}+448451848 x^{14}+196808960 x^{13}+1186429342 x^{12}+196808960 x^{11}\nn\\
  &\quad+448451848 x^{10}-168688780 x^9-240496596 x^8-31412980 x^7-18036936 x^6+3106950 x^5\nn\\
  &\quad+4669125 x^4+185850 x^3+119808 x^2-15360+840 \left(x^2-1\right)^2 \left(525 x^{14}+2498 x^{13}+5430 x^{12}\right.\nn\\
  &\quad+6324 x^{11}-5888 x^{10}-30882 x^9-60483 x^8-70568 x^7-60483 x^6-30882 x^5-5888 x^4+6324 x^3\nn\\
  &\quad\left.+5430 x^2+2498 x+525\right) x^3 \log (2)\,,\nn\\
  h_{21}(x) &= -525 x^{14}+2498 x^{13}-5430 x^{12}+6324 x^{11}+5888 x^{10}-30882 x^9+60483 x^8-70568 x^7+60483 x^6\nn\\
  &\quad-30882 x^5+5888 x^4+6324 x^3-5430 x^2+2498 x-525\,,\nn\\
  h_{22}(x) &= -2273 x^{12}-6234 x^{10}+202833 x^8+545236 x^6+202833 x^4-6234 x^2-2273\,,\nn\\
  h_{23}(x) &= \left(x^2+1\right) \left(525 x^{12}-4096 x^{11}-1110 x^{10}-8192 x^9-285 x^8+69632 x^7-4404 x^6+69632 x^5-285 x^4\right.\nn\\
  &\quad\left.-8192 x^3-1110 x^2-4096 x+525\right)\,,\nn\\
  h_{24}(x) &= -6015 x^{18}+20963 x^{16}+380928 x^{15}-9540 x^{14}+16384 x^{13}-58956 x^{12}-4055040 x^{11}+110390 x^{10}\nn\\
  &\quad-4214784 x^9-64734 x^8-61440 x^7-6756 x^6+466944 x^5+19028 x^4-3855 x^2-4096 x-525\,,\nn\\
  h_{25}(x) &= -512 \left(9 x^{18}-63 x^{14}+263 x^{12}-1509 x^{10}-2697 x^8-5 x^6+357 x^4-3\right)\,,\nn\\
  h_{26}(x) &= -512 \left(7 x^{18}-675 x^{14}+189 x^{12}+4833 x^{10}+3159 x^8-5 x^6-213 x^4+1\right)\,,\nn\\
  h_{27}(x) &= 1024 \left(3 x^{12}+9 x^{10}-207 x^8-532 x^6-189 x^4+3 x^2+1\right)\,,\nn\\
  h_{28}(x) &= 6015 x^{18}-20963 x^{16}+380928 x^{15}+9540 x^{14}+16384 x^{13}+58956 x^{12}-4055040 x^{11}-110390 x^{10}\nn\\
  &\quad-4214784 x^9+64734 x^8-61440 x^7+6756 x^6+466944 x^5-19028 x^4+3855 x^2-4096 x+525\,,\nn\\
  h_{29}(x) &= 1823 x^{12}+6054 x^{10}-202671 x^8-544300 x^6-202671 x^4+6054 x^2+1823\,,\nn\\
  h_{30}(x) &= -\left(\left(x^2+1\right) \left(525 x^{12}+4096 x^{11}-1110 x^{10}+8192 x^9-285 x^8-69632 x^7-4404 x^6-69632 x^5\right.\right.\nn\\
  &\quad\left.\left.-285 x^4+8192 x^3-1110 x^2+4096 x+525\right)\right)\,,\nn\\
  h_{31}(x) &= 525 x^{14}+2498 x^{13}+5430 x^{12}+6324 x^{11}-5888 x^{10}-30882 x^9-60483 x^8-70568 x^7-60483 x^6\nn\\
  &\quad-30882 x^5-5888 x^4+6324 x^3+5430 x^2+2498 x+525\nn
\end{align}

\begin{align}
  d_1(x) &= -\nu ^2 \left(\nu ^2 \left(385 x^{14}+590 x^{13}-5355 x^{12}+9960 x^{11}-12517 x^{10}+8802 x^9-9137 x^8+6352 x^7-9137 x^6\right.\right.\nn\\
  &\quad\left.+8802 x^5-12517 x^4+9960 x^3-5355 x^2+590 x+385\right) (x-1)^6+8 \nu  x \left(175 x^{14}+650 x^{13}-970 x^{12}\right.\nn\\
  &\quad+2972 x^{11}-2012 x^{10}+3414 x^9-905 x^8+3080 x^7-905 x^6+3414 x^5-2012 x^4+2972 x^3-970 x^2\nn\\
  &\quad\left.+650 x+175\right) (x-1)^4+32 x^2 \left(35 x^{16}+105 x^{15}-310 x^{14}+725 x^{13}-995 x^{12}+1331 x^{11}-1150 x^{10}\right.\nn\\
  &\quad\left.\left.+1103 x^9-920 x^8+1103 x^7-1150 x^6+1331 x^5-995 x^4+725 x^3-310 x^2+105 x+35\right)\right)\,,\nn\\
  d_2(x) &= 12 \left(-\nu  \left(x^{20}+14 x^{19}+10 x^{18}-165 x^{16}-1350 x^{15}-772 x^{14}+378 x^{13}+2744 x^{12}+9666 x^{11}+6572 x^{10}\right.\right.\nn\\
  &\quad\left.+6318 x^9+2756 x^8-10 x^7-700 x^6-426 x^5-217 x^4+10 x^2+2 x+1\right)+7 x^{19}-675 x^{15}+189 x^{13}\nn\\
  &\quad\left.+4833 x^{11}+3159 x^9-5 x^7-213 x^5+x\right)\,,\nn\\
  d_3(x) &= -3 \left(\nu  (x-1)^2 \left(525 x^{12}-3046 x^{11}+4828 x^{10}+414 x^9-14381 x^8+32264 x^7-37368 x^6+32264 x^5\right.\right.\nn\\
  &\quad\left.-14381 x^4+414 x^3+4828 x^2-3046 x+525\right)-525 x^{14}+2498 x^{13}-5430 x^{12}+6324 x^{11}+5888 x^{10}\nn\\
  &\quad\left.-30882 x^9+60483 x^8-70568 x^7+60483 x^6-30882 x^5+5888 x^4+6324 x^3-5430 x^2+2498 x-525\right)\,,\nn\\
  d_4(x) &= 3 \left(\nu  \left(1575 x^{14}+5896 x^{13}+10275 x^{12}+13008 x^{11}-13171 x^{10}-62088 x^9-125655 x^8-143008 x^7\right.\right.\nn\\
  &\quad\left.-125655 x^6-62088 x^5-13171 x^4+13008 x^3+10275 x^2+5896 x+1575\right)-525 x^{14}-2498 x^{13}-5430 x^{12}\nn\\
  &\quad-6324 x^{11}+5888 x^{10}+30882 x^9+60483 x^8+70568 x^7+60483 x^6+30882 x^5+5888 x^4\nn\\
  &\quad\left.-6324 x^3-5430 x^2-2498 x-525\right)\,,\nn\\
  d_5(x) &= \Gamma  \nu ^2 \left(\nu  (x-1)^4+2 \left(x^3+x\right)\right)^2 \left(\nu  \left(10752 x^{38}+170752 x^{37}+577878 x^{36}+3053105 x^{35}+984228 x^{34}\right.\right.\nn\\
  &\quad-23188558 x^{33}-102246102 x^{32}-395889516 x^{31}-760723306 x^{30}-1705980266 x^{29}-1596565222 x^{28}\nn\\
  &\quad-2276395052 x^{27}+3231831134 x^{26}+5854097682 x^{25}+25834221154 x^{24}+31361563244 x^{23}\nn\\
  &\quad+68452047454 x^{22}+66134059430 x^{21}+106203539470 x^{20}+83073443318 x^{19}+106203539470 x^{18}\nn\\
  &\quad+66134059430 x^{17}+68452047454 x^{16}+31361563244 x^{15}+25834221154 x^{14}+5854097682 x^{13}\nn\\
  &\quad+3231831134 x^{12}-2276395052 x^{11}-1596565222 x^{10}-1705980266 x^9-760723306 x^8-395889516 x^7\nn\\
  &\quad\left.-102246102 x^6-23188558 x^5+984228 x^4+3053105 x^3+577878 x^2+170752 x+10752\right)-7 \left(768 x^{38}\right.\nn\\
  &\quad+41277 x^{36}+70302 x^{34}-7303293 x^{32}-54337379 x^{30}-114040373 x^{28}+230845081 x^{26}+1845301511 x^{24}\nn\\
  &\quad+4889431961 x^{22}+7585967105 x^{20}+7585967105 x^{18}+4889431961 x^{16}+1845301511 x^{14}+230845081 x^{12}\nn\\
  &\quad\left.\left.-114040373 x^{10}-54337379 x^8-7303293 x^6+70302 x^4+41277 x^2+768\right)\right)\nn\\
  &\quad-280 \nu ^2 x \left(x^2-1\right)^2 \left(x^2+1\right)^9 \left(\nu ^3 \left(385 x^{17}-720 x^{16}-2935 x^{15}+11280 x^{14}-16357 x^{13}+11668 x^{12}\right.\right.\nn\\
  &\quad-869 x^{11}-1616 x^{10}+1727 x^9+4492 x^8-6553 x^7+8752 x^6+597 x^5-10868 x^4+15285 x^3-10160 x^2\nn\\
  &\quad\left.+3600 x-540\right) (x-1)^6+\nu ^2 x \left(1785 x^{19}-3930 x^{18}-9230 x^{17}+55878 x^{16}-122324 x^{15}+154718 x^{14}\right.\nn\\
  &\quad-114554 x^{13}+40958 x^{12}+6514 x^{11}+5342 x^{10}-40506 x^9+77518 x^8-62380 x^7-9182 x^6+102394 x^5\nn\\
  &\quad\left.-142718 x^4+115733 x^3-59268 x^2+18120 x-2580\right) (x-1)^2+8 \nu  x^2 \left(315 x^{19}-375 x^{18}-1955 x^{17}\right.\nn\\
  &\quad+7701 x^{16}-14693 x^{15}+17822 x^{14}-13841 x^{13}+5750 x^{12}+109 x^{11}+1148 x^{10}-5121 x^9+9556 x^8\nn\\
  &\quad\left.-7183 x^7-1382 x^6+11713 x^5-16334 x^4+14012 x^3-8061 x^2+2820 x-465\right)+16 x^3 \left(70 x^{17}+105 x^{16}\right.\nn\\
  &\quad-330 x^{15}+620 x^{14}-747 x^{13}+606 x^{12}-337 x^{11}-228 x^{10}-216 x^9-676 x^7+228 x^6-27 x^5-606 x^4\\
  &\quad\left.\left.+623 x^3-620 x^2+360 x-105\right)\right)\,,\nn\\
  d_6(x) &= \nu ^2 \left(\nu ^2 \left(385 x^{14}+590 x^{13}-5355 x^{12}+9960 x^{11}-12517 x^{10}+8802 x^9-9137 x^8+6352 x^7-9137 x^6\right.\right.\nn\\
  &\quad\left.+8802 x^5-12517 x^4+9960 x^3-5355 x^2+590 x+385\right) (x-1)^6+8 \nu  x \left(175 x^{14}+650 x^{13}-970 x^{12}\right.\nn\\
  &\quad+2972 x^{11}-2012 x^{10}+3414 x^9-905 x^8+3080 x^7-905 x^6+3414 x^5-2012 x^4+2972 x^3-970 x^2+650 x\nn\\
  &\quad\left.+175\right) (x-1)^4+32 x^2 \left(35 x^{16}+105 x^{15}-310 x^{14}+725 x^{13}-995 x^{12}+1331 x^{11}-1150 x^{10}+1103 x^9\right.\nn\\
  &\quad\left.\left.-920 x^8+1103 x^7-1150 x^6+1331 x^5-995 x^4+725 x^3-310 x^2+105 x+35\right)\right)\,,\nn\\
  d_7(x) &= -3 \left(\nu  \left(525 x^{14}-7292 x^{13}-585 x^{12}-24216 x^{11}-54643 x^{10}+810684 x^9-90705 x^8\right.\right.\nn\\
  &\quad\left.+2177200 x^7-90705 x^6+810684 x^5-54643 x^4-24216 x^3-585 x^2-7292 x+525\right)+2 x \left(1823 x^{12}\right.\nn\\
  &\quad\left.\left.+6054 x^{10}-202671 x^8-544300 x^6-202671 x^4+6054 x^2+1823\right)\right)\,,\nn\\
  d_8(x) &= \nu  \left(5775 x^{14}-11884 x^{13}+65745 x^{12}-47352 x^{11}-122509 x^{10}+932268 x^9-807123 x^8+2448240 x^7\right.\nn\\
  &\quad\left.-807123 x^6+932268 x^5-122509 x^4-47352 x^3+65745 x^2-11884 x+5775\right)-3150 x^{14}+8642 x^{13}\nn\\
  &\quad-32580 x^{12}+24756 x^{11}+35328 x^{10}-467106 x^9+362898 x^8-1229736 x^7+362898 x^6-467106 x^5\nn\\
  &\quad+35328 x^4+24756 x^3-32580 x^2+8642 x-3150\,,\nn\\
  d_9(x) &= \Gamma  \nu ^2 \left(\nu  (x-1)^4+2 \left(x^3+x\right)\right)^2 \left(\nu  \left(188160 x^{34}-39829184 x^{33}-68749940 x^{32}-358154923 x^{31}\right.\right.\nn\\
  &\quad+923064940 x^{30}+6460709986 x^{29}+17452593620 x^{28}+58553910415 x^{27}+69306173300 x^{26}\nn\\
  &\quad+163660060004 x^{25}+26514969180 x^{24}+64744381653 x^{23}-531452366340 x^{22}-695910106626 x^{21}\nn\\
  &\quad-1819656084540 x^{20}-1960277378745 x^{19}-3057751871580 x^{18}-2626177239560 x^{17}-3057751871580 x^{16}\nn\\
  &\quad-1960277378745 x^{15}-1819656084540 x^{14}-695910106626 x^{13}-531452366340 x^{12}+64744381653 x^{11}\nn\\
  &\quad+26514969180 x^{10}+163660060004 x^9+69306173300 x^8+58553910415 x^7+17452593620 x^6\nn\\
  &\quad\left.+6460709986 x^5+923064940 x^4-358154923 x^3-68749940 x^2-39829184 x+188160\right)-490 \left(192 x^{34}\right.\nn\\
  &\quad-70153 x^{32}+941903 x^{30}+17808769 x^{28}+70720585 x^{26}+27056091 x^{24}-542298333 x^{22}-1856791923 x^{20}\nn\\
  &\quad-3120154971 x^{18}-3120154971 x^{16}-1856791923 x^{14}-542298333 x^{12}+27056091 x^{10}+70720585 x^8\nn\\
  &\quad\left.\left.+17808769 x^6+941903 x^4-70153 x^2+192\right)\right)-9800 \nu ^2 (x-1)^2 \left(x^2+1\right)^8 \left(\nu ^3 \left(1155 x^{18}-6738 x^{17}\right.\right.\nn\\
  &\quad+18579 x^{16}-51312 x^{15}+168696 x^{14}-389296 x^{13}+490760 x^{12}-309040 x^{11}-86294 x^{10}+240964 x^9\nn\\
  &\quad\left.-86294 x^8-309040 x^7+490760 x^6-389296 x^5+168696 x^4-51312 x^3+18579 x^2-6738 x+1155\right) (x-1)^6\nn\\
  &\quad+\nu ^2 x \left(5355 x^{20}-33756 x^{19}+108102 x^{18}-338700 x^{17}+1074299 x^{16}-2555024 x^{15}+4170464 x^{14}\right.\nn\\
  &\quad-4614000 x^{13}+2831578 x^{12}+131816 x^{11}-1707724 x^{10}+131816 x^9+2831578 x^8-4614000 x^7+4170464 x^6\nn\\
  &\quad\left.-2555024 x^5+1074299 x^4-338700 x^3+108102 x^2-33756 x+5355\right) (x-1)^2+4 \nu  x^2 \left(1890 x^{20}-10191 x^{19}\right.\nn\\
  &\quad+28941 x^{18}-100995 x^{17}+326992 x^{16}-683404 x^{15}+981376 x^{14}-1053108 x^{13}+701390 x^{12}-4718 x^{11}\nn\\
  &\quad-413210 x^{10}-4718 x^9+701390 x^8-1053108 x^7+981376 x^6-683404 x^5+326992 x^4-100995 x^3\nn\\
  &\quad\left.+28941 x^2-10191 x+1890\right)+8 x^3 \left(420 x^{18}-1167 x^{17}+2355 x^{16}-14184 x^{15}+34723 x^{14}\right.\nn\\
  &\quad-35944 x^{13}+42525 x^{12}-38792 x^{11}+4457 x^{10}+41934 x^9+4457 x^8-38792 x^7+42525 x^6-35944 x^5\nn\\
  &\quad\left.\left.+34723 x^4-14184 x^3+2355 x^2-1167 x+420\right)\right)\nn
\end{align}

\endgroup
\end{widetext} 
\end{document}